\documentclass[a4paper,11pt]{article}

\usepackage[american]{babel}
\usepackage[intlimits,leqno]{amsmath}
\usepackage{amssymb}
\usepackage{epsfig}
\usepackage{latexsym}

\newtheorem{rema}{Remark}
\newcommand{\A}{\mathcal{A}}
\newcommand\R{{\mathbb{R}}}
\newcommand\N{{\mathbb{N}}}
\newcommand\Z{{\mathbb{Z}}}
\newcommand{\x}{\mathbf{x}}

\newcommand\e{{\mathbf{e}}}
\newcommand\vv{{\mathbf{v}}}
\newcommand\vj{{\mathbf{j}}}

\newcommand\vO{{\mathbf{0}}}
\newcommand\vA{{\mathbf{A}}}
\newcommand\vB{{\mathbf{B}}}
\newcommand\vE{{\mathbf{E}}}
\newcommand\vf{{\mathbf{f}}}
\newcommand\vS{{\mathbf{S}}}
\renewcommand{\d}{{\rm d}}

\begin{document}

\title{On the causality of electrodynamics
and the Aharonov-Bohm effect}

\author{ Richard Kowar\\
Department of Mathematics, University of Innsbruck, \\
Technikerstrasse 21a/2, A-6020, Innsbruck, Austria}

\maketitle

\begin{abstract}
This paper presents a \emph{non-instant field model} for electrodynamics 
that permits a causal explanation of the \emph{Aharonov-Bohm effect} and 
a \emph{covariant quantization} of the respective Maxwell equations via 
the \emph{Gupta-Bleuler method}.  
Our model satisfies the following \emph{correspondence principle}: if 
$A^\mu$, $\vE$, $\vB$ denote the four potential, the electric field and 
the magnetic field of the non-instant field model, then the respective 
classical quantities are $\A[A^\mu]$, $\A[\vE]$, $\A[\vB]$, where $\A$ is 
a covariant time averaging operator. Here $\A[A^\mu]$ is interpreted as 
the best possible measurement of the four potential $A^\mu$. Although the 
Lorentz condition is not satisfied for $A^\mu$, it is satisfied for 
$\A[A^\mu]$. The latter fact means that the Lorentz condition does not 
hold for the quantized field but for its expectation value (cf. 
\emph{Gupta-Bleuler method} of quantization). Finally, we 
derive the energy conservation law of our field model and show that the 
field energy is quantized. 
\end{abstract}

\vspace{2pc}
\maketitle

\section{Introduction}
\label{sec-intro}

The goal of this paper is to present and discuss a non-instant field model of charged 
particles that has stronger causality properties than the classical model and can be 
quantized. A highlight of this paper is the causal explanation of the 
\emph{Aharonov-Bohm effect}. It is best to start with a shortly review of electrodynamics 
and discuss the differences and relations to our model without carrying out the details.

The electromagnetic field concept in classical field theory is an instant one and in 
this sense causality is violated. For example, the four potential $A^\mu_{cl}=(\phi_{cl},\vA_{cl})$ 
of a charged particle in its restframe $S'$ is modeled by (cf.~\cite{Fey89II,Jac99})
$$
  \phi'_{cl} = \frac{Q_0}{|\x'|}
\qquad\mbox{and}\qquad
   \vA'_{cl} = 0\,,
$$
where a prime indicates a quantity observed in the restframe $S'$ of the charge. The charge 
$Q_0$ is an invariant and therefore is not primed. We see at once that the Coulomb condition 
$\nabla'\cdot \vA'_{cl}=0$ as well as the Lorentz condition 
$\partial_\mu'\,{A'}^\mu_{cl}=\frac{1}{c}\,\frac{\partial \phi'_{cl}}{\partial t'}+\nabla'\cdot \vA'_{cl}= 0$ 
are satisfied. Since the latter condition is covariant (cf.~\cite{Wei72,Jac99}), it holds in all 
reference frames. In contrast to this model, a non-instant field model must have a time-dependent 
scalar potential $\phi'=\phi'(\x',t')$ and if $\vA' = 0$ is assumed, then the Lorentz condition  
cannot be satisfied. Consequently, for the non-instant field model presented in this paper
$$
      \mbox{$q:= \partial_\mu A^\mu\quad$ is a non-vanishing \emph{invariant}}
$$
and the respective field theory depends not only on the electric and magnetic fields $\vE$ and 
$\vB$ but also on the \emph{invariant} $q$. As a consequence, a \emph{Lorentz gauge transformation} 
of the non-instant four potential $A^\mu$ is prohibited. Since gauge transformations are used all 
the time in physics (cf.~\cite{Fey89II,LanLif89,Jac99,Wei72,Wei95,ManSha10,Sch93,Sch97,Zee03}), 
this statement seems to be absurd. 
But on the other hand, gauge theories might be considered as ``unsatisfactory'' (cf. Chapter 
III.4 in~\cite{Zee03}). 
It turns out that this is not a problem, since our model satisfies the following 
\emph{correspondence principle}.

The respective classical quantities of electrodynamics can be obtained from the non-instant field 
model via a covariant time averaging operator $\A$. If $\phi'={A'}^0$, $\vE'$ and $\vB'$ denote 
the scalar potential, the electric field and the magnetic field of a charged particle in 
its restframe $S'$, then it follows 
$$
  \A[\phi'] = \phi'_{cl}\,,\qquad
  \A[\vE'] = \vE'_{cl}\,,\qquad
  \A[\vB'] = \vB'_{cl}
\qquad\mbox{and}\qquad
   \partial_\mu\,\A[A^\mu] = 0\,,
$$
where $\phi'_{cl}$, $\vE'_{cl}$ and $\vB'_{cl}$ denote the respective quantities from classical 
electrodynamics. 
The last condition means that the Lorentz condition is satisfied for the time averaged non-instant 
four potential. Here the quantity $\A[A^\mu]$ is interpreted as the best possible measurement  
of the four potential $A^\mu$. Because the time averaging interval $\tau'$ is extremely small, it 
appears as if the measurement process was performed at a time instant (in $S'$). 

Because the classical quantities $\A[\phi']$ and $\A[\vE']$ are time independent, it seems as if 
the electromagnetic field fills out the space instantaneously. 
On the one hand, calculation with time averaged quantities is very convenient (because we have 
gauge invariance and we got used to it), but on the other hand, the ``time averaged picture'' 
does not always allow to associate a cause to an  effect. 
For example, the non-instant field model predicts that a solenoid sends out an electromagnetic 
wave of extreme high frequency $2\pi/\tau$ such that the \emph{Aharonov-Bohm effect} 
(cf.~\cite{AhaBoh59,Fey89II,Sch93,Gre93,Zee03,BerSch93}) does not violate causality. However, 
if we go over to the "time averaged picture" causality seems to be violated, since then 
$$
  \vE_{cl} \equiv \A[\vE]=\vO \qquad\mbox{and}\qquad  \vB_{cl} \equiv \A[\vB]=\vO 
  \qquad \mbox{outside the solenoid.}
$$
The \emph{correspondence principle} can be summarized as follows:
\begin{equation*}
\begin{aligned}
     &\mbox{relevant \& causal quantity $A^\mu$}
\quad\stackrel{\A}{\longrightarrow}\quad
     \mbox{measurable quantity $A^\mu_{cl}$} \\
     &\qquad\qquad   \mbox{(unknown)}
      \qquad\qquad\quad\, \mbox{(filter)}
      \qquad\quad\;      \mbox{(known)}
\end{aligned}
\end{equation*}
For modeling the left side of this mapping there might be only one reason/justification, 
to find the causes of a measurable effects.

Although this paper is mainly concerned with classical field theory I would like to say a view words 
about quantum electrodynamics. Since Maxwell's equations of the non-instant four potential read as 
follows
\begin{equation*}
\begin{aligned}
  &\partial_\nu \,F^{\mu\,\nu} - \partial^\mu \,(\partial_\nu\,A^\nu)
   =  - 4\,\pi\,\frac{j^\mu}{c}
   \qquad\quad (\mu=0,\,1,2\,,\,3)\,,
\end{aligned}
\end{equation*}
where $j^\mu$ is the four current and $F_j^{\mu\,\nu} = \partial^\mu A_j^\nu - \partial^\nu A_j^\mu$ is the 
electromagnetic field tensor (cf. Sections~\ref{sec-fields} and~\ref{sec-max}), these equations have 
the same form as the modified equations used for the quantization  of  the electromagnetic field via 
the \emph{Gupta-Bleuler method} (cf. Subsection 5.1 in~\cite{ManSha10}). That the Lorentz condition 
is only satisfied for $\A[A^\mu]$ means that the Lorentz condition is satisfied for the expectation 
value of the quantized four potential, which is (and can be) assumed in quantum electrodynamics. 
In contrast to quantum electrodynamics we can use the previous equations without assuming the Lorentz 
condition for the four potential which cannot be quantized. \\

This paper is organized as follows. Firstly, we introduce our non-instant field model in 
Section~\ref{sec-fields}. Secondly, we derive the respective Maxwell equatiosn in covariant and 
non-covariant form in Section~\ref{sec-max}. Afterwards, in Section~\ref{sec-Cp1}, the correspondence 
principle is introduced and discussed for the case of a charged particle and a solenoid. 
Section~\ref{sec-Cp1} is concluded with the causal explanation of the Aharonov-Bohm effect. 
The conservation law of energy and the quantization of the field energy is derived in Section~\ref{sec-energy}. 
The paper is concluded with the section "Results" containing the subsection "Conclusions".

\section{Non-instant field model of charged particles}
\label{sec-fields}

\subsection{The four potential $A^\mu$}
\label{sec-defA}

In the following we define a four potential $A^\mu=(\phi,\vA)$ of a
particle with charge $Q_0$ observed from an inertial frame $S_I$.
We do not assume that the restframe $S'$ of the particle is an inertial
system. In its restframe $S'$ the particle is created at $t'=0$ and
destroyed at $t'=N\,\tau'$. In this paper 
$$
\mbox{$\tau'>0$ is the same constant for each particle in its restframe.}
$$
The field $A^\mu$ is a superposition of fields $A_j^\mu$ that are generated 
at the time instants $\tau_j'=j\,\tau'$ in $S'$ with 
$j\in J:=\{0,\,1,\,\ldots,\,N-1\}$. It is convenient to introduce $S'_j$ 
as the inertial systems such that its origin and axes coincide with those 
of system $S'$ at time $t=\tau_j$ in $S_I$. \\
i) We define the field ${A'}_j^\mu$ in the inertial system $S'_j$ by
\begin{equation*}
\begin{aligned}
     {A'}_j^\mu = (\phi_j',\vA_j')
  \qquad\qquad  j\in J \,,
\end{aligned}
\end{equation*}
with 
\begin{equation}\label{defA}
\begin{aligned}
      \phi_j'(\x',t')
          = Q_0\,
         \frac{f'_0(\omega'\,(t'-\tau_j')-k'\,|\x'|)}{|\x'|}
  \qquad\mbox{and}\qquad
      \vA_j' = \vO\,,
\end{aligned}
\end{equation}
where
\begin{equation}\label{defomegak}
  \mbox{$Q_0$ is an invariant,}\qquad\qquad
   \omega':=\frac{2\,\pi}{\tau'} \,,
\qquad\qquad
   k':=\frac{\omega'}{c}\,,
\end{equation}
the \emph{smooth function} $t'\in\R\mapsto f_0(\omega'\,t')\in\R$ is positive,
symmetric with respect to $t'=\tau'/2$ and has support
in $[0,\tau']$,\footnote{We use the Gaussian unit system in this paper, i.e. 
$\epsilon_0=\mu_0=1$. Both right hand sides in~(\ref{deff0}) correspond to 
$1/[\epsilon_0]\equiv 1$.} 
\begin{equation}\label{deff0}
\begin{aligned}
     \,\int_\R \frac{f_0'(r)}{2\,\pi}\,\d r = 1
\qquad\mbox{and}\qquad
   4\,\pi\,\alpha\,\int_\R \left[\frac{\d {f'_0(r)}}{\d r}\right]^2\,\d r =  1\,.
\end{aligned}
\end{equation}
Here $c$, $e$ and $\alpha=\frac{e^2}{c\,\hbar}$ denote the \emph{speed of light}, 
the \emph{charge of the electron} and the \emph{fine structure constant}, 
respectively. The last two conditions can be satisfied for an appropriate
choice of $f_0'$. \\
ii) Now we consider the situation from an inertial frame $S_I$ satisfying
$t=0$ if $t'=0$ in $S'$. The four potential $(\phi_j,\vA_j)$ is defined
in $S_I$ by (cf.~\cite{Fey89II,Jac99,LanLif89,Wei72}) 
\begin{equation*}
\begin{aligned}
  \phi_j^2(\x,t) - |\vA_j(\x,t)|^2
        = \phi^{'2}_j(\x',t') \,,
\end{aligned}
\end{equation*}
where $\x'(\x,t)$ and $t'(\x,t)$ are the coordinates of $S'_j$
with respect to system $S_I$. More suggestive
\begin{equation*}
\begin{aligned}
  \phi_j[\vv_j]^2 - |\vA_j[\vv_j]|^2
        = \phi'_j[\vO]^2 \,,
\end{aligned}
\end{equation*}
where $\vv_j$ denotes the velocity of $S'_j$ with respect to $S_I$ at
$t=\tau_j$. The total four potential is given by
\begin{equation}\label{Atotal}
         A^\mu = \sum_{j\in J} A_j^\mu[\vv_j] \,.
\end{equation}

\begin{rema}
We note that $Q_0$ and $q'=\partial_\mu {A'}^\mu
=\frac{1}{c}\,\frac{\partial \phi'}{\partial t'}$ (under the assumption 
$f'_0(-\infty)=0$) determine the field model of a particle created at 
time $t=0$ uniquely, since
\begin{equation*}
\begin{aligned}
   &\mbox{$\tau'$ is the smallest positive real number such that} \\
   &\qquad\qquad \int_0^{\tau'} q'(\x',r)\,\d r = 0\,
\end{aligned}
\end{equation*}
and
\begin{equation*}
\begin{aligned}
   f'_0(r') - f'_0(-\infty)
      = -\frac{c^2\,\tau'\,r'}{2\,\pi\,Q_0}\,
         \int_{-\infty}^0 q'\left( - \frac{c\,\tau'\,r'}{2\,\pi},0,0,t\right)\,\d t\,.
\end{aligned}
\end{equation*}
Here the assumption $f'_0(-\infty)=0$ is considered as a causality assumption. 
If the particle is created at time $t=0$, then $f'_0(t)=0$ for $t<0$ is required 
by causality.  

For the classical model in electrodynamics with Lorentz gauge, we have $q=0$ and thus 
\begin{equation*}
\begin{aligned}
   f'_0(r') - f'_0(-\infty)  = 0  
\qquad\mbox{with}\qquad 
   f'_0(-\infty) = 1\,,
\end{aligned}
\end{equation*}
i.e. causality is not satisfied. 
\end{rema}

\subsection{The electromagnetic field and its tensor}
\label{sec-defEB}

As in classical field theory we define the electromagnetic field of a
charged particle via its four potential (cf.~\cite{Fey89II,LanLif89,Jac99,Wei72}).
In the inertial frame $S_I$, we define the total electric and magnetic fields
by $\vE := \sum_{j\in J} \vE_j$ and $\vB := \sum_{j\in J} \vB_j$
with
\begin{equation}\label{defEjBj}
\begin{aligned}
   &\vE_j
          = -\nabla \phi_j
            - \frac{1}{c}\,\frac{\partial \vA_j}{\partial t}
         \qquad\mbox{and}\qquad
   \vB_j
      = \nabla \times \vA_j \,.
\end{aligned}
\end{equation}
The \emph{electromagnetic field tensor} is defined by
$F^{\mu\,\nu} = \sum_{j\in J} F_j^{\mu\,\nu}$ with
\begin{equation}\label{defftrensor01}
\begin{aligned}
  F_j^{\mu\,\nu}
       = \partial^\mu A_j^\nu - \partial^\nu A_j^\mu\,,
\end{aligned}
\end{equation}
where $\partial_\nu
     := \left(\frac{1}{c}\,\frac{\partial}{\partial t},\nabla\right)$
and $\partial^\nu
     := \left(\frac{1}{c}\,\frac{\partial}{\partial t},-\nabla\right)$.
Because $A^\mu$ is a four vector, $F_j^{\mu\,\nu}$ is a tensor and
${(F_j)}_{\mu\,\nu}\,(F_j)^{\mu\,\nu}$,
$\epsilon^{\mu\,\nu\,\alpha\,\beta}\,{(F_j)}_{\mu\,\nu}\,{(F_j)}_{\alpha\,\beta}$
are invariants (cf.~\cite{LanLif89,Wei72,GreRaf92}).\footnote{Strictly speaking
the second quantity is a pseudoscalar.}

\subsection{Electromagnetic four force}
\label{sec-deff}

As usual, the \emph{electromagnetic four force} on a particle
with charge $Q$ and velocity $\vv$ is given by (cf.~\cite{Fey89II,LanLif89})
$$
     \vf^\mu:=Q\,F^{\mu\,\nu}\,u_\nu\,,
$$
where $u^\mu$ denotes the \emph{four velocity} $u^\mu := (1-v^2/c^2)^{-1/2}\,(c,\vv)$. \\

\begin{rema}\label{rema-force}
Consider a charged particle in its restframe. Since its field $A^\mu$ oscillates 
at a fixed point in space, the same is true for the electromagnetic field tensor 
$F^{\mu\,\nu}$ and hence the four force onto another particle "oscillates" too. 
But this means that the trajectory of the second particle is quite "chaotic" at 
a very small scale. However, the respective time averaged trajectory is less 
chaotic and smoother, since we integrate over time. But this fact will not be 
investigated closer in this paper. We do not need the concept of force in this 
paper. 
\end{rema}

\section{Maxwell's equations for the non-instant field model}
\label{sec-max}

We now derive the Maxwell equations for the non-instant field model as 
defined in Section~\ref{sec-fields}. The most striking result will be 
- without assuming the Lorentz condition - that these equations
(cf.~(\ref{Maxwell01a}) with~(\ref{defftrensor01})) are equivalent to
the form of the Maxwell equations used in~\cite{ManSha10} for the
quantization of the electromagnetic field via the
\emph{Gupta-Bleuler method}.

\subsubsection*{Covariant form for $A^\mu$}

From analysis, it is known that $A_j^{'\mu}=(\phi_j',\vA_j')$
defined as in Subsection~\ref{sec-defA} satisfies the wave
equations (cf.~\cite{Tre95,DauLio92_5,Hoe03})
\begin{equation}\label{eqphi}
\begin{aligned}
 \nabla^{'2} \phi_j'
        - \frac{1}{c^2}\,\frac{\partial^2 \phi_j'}{\partial t^{'2}}
    = - 4\,\pi\,\rho_j' \,
\qquad\mbox{and}  \qquad
   \nabla^{'2} \vA_j'
       - \frac{1}{c^2}\,\frac{\partial^2 \vA_j'}{\partial t^{'2}}
          = - 4\,\pi\,\frac{\vj_j'}{c}\,,
\end{aligned}
\end{equation}
where
\begin{equation}\label{defj}
\begin{aligned}
      \rho_j'
    = Q_0\,f'_0(\omega'\,(t'-\tau'_j))\,\delta(\x')\,
\qquad  \mbox{and}\qquad
      \vj_j' = \vO\,.
\end{aligned}
\end{equation}
We define $j_j^\mu$ as the four vector satisfying
$j_j^{'\mu}=(c\,\rho_j',\vj_j')$ in $S'_j$ and $j^\mu$ as the total
\emph{four current} $j^\mu:=\sum_{j\in J}j_j^\mu$.
Because $A^\mu$ (cf.~(\ref{Atotal})) and $j^\mu$ are four vectors, it follows
from~(\ref{eqphi}) and~(\ref{defj}) that (cf.~\cite{Wei72}).
\begin{equation}\label{Maxwell01}
\begin{aligned}
   \nabla^2 A^\mu
       - \frac{1}{c^2}\,\frac{\partial^2 A^\mu}{\partial t^2}
          = - 4\,\pi\,\frac{j^\mu}{c}
\qquad\quad (\mu=0,\,1,2\,,\,3)\,.
\end{aligned}
\end{equation}
Formally, these equations are equivalent to the classical equations of Maxwell 
if $\partial_\mu A_j^\mu=0$, i.e. the Lorentz condition is satisfied
(cf. Chapter~25 in~\cite{Fey89II}).
However, $\phi_j'$ is time dependent for our non-instant field model, i.e.
$$
        \partial_\mu {A}_j^\mu
           = \partial'_\mu {A'}_j^\mu
           = \frac{1}{c}\,\frac{\partial \phi_j'}{\partial t'}
          \not=0
$$
and thus each
\begin{equation}\label{defqj}
   q_j := \partial_\mu A_j^\mu \qquad
          \mbox{is an invariant that is not identical zero.}
\end{equation}
Consequently, a \emph{Lorentz gauge transformation} is prohibited for the
non-instant field model presented in this paper. \\

\noindent
With the electromagnetic field tensor $F^{\mu\,\nu}$ defined as
in~(\ref{defftrensor01}) and the scalar
\begin{equation}\label{defq}
\begin{aligned}
      q := \sum_{j\in J} q_j = \partial^\nu A^\mu \,,
\end{aligned}
\end{equation}
the equations of Maxwell~(\ref{Maxwell01}) read as follows
(cf.~\cite{Fey89II,LanLif89,Wei72})
\begin{equation}\label{Maxwell01a}
\begin{aligned}
  \partial_\nu \,F^{\mu\,\nu} - \partial^\mu \,q
   =  - 4\,\pi\,\frac{j^\mu}{c}
   \qquad\quad (\mu=0,\,1,2\,,\,3)\,.
\end{aligned}
\end{equation}
These are the Maxwell equations used in~\cite{ManSha10} for the
quantization of the electromagnetic field via the
\emph{Gupta-Bleuler method}.

\subsubsection*{Non-covariant form}

From~(\ref{defEjBj}),~(\ref{defqj}),~(\ref{Maxwell01}) and the identities
$$
    \nabla\times (\nabla \phi) = \vO
\qquad \mbox{and}\qquad
   \nabla\times (\nabla\times \vA)
        = \nabla(\nabla \cdot \vA) -\nabla^2 \vA\,,
$$
it follows that
\begin{equation}\label{Maxwell01b}
\begin{aligned}
    &\nabla \cdot \vE
        = 4\,\pi\,\rho
         - \frac{1}{c}\,\frac{\partial q}{\partial t}\,,
\qquad\qquad \qquad\qquad
    \nabla\times \vE
        = - \frac{1}{c}\,\frac{\partial \vB}{\partial t}\,, \\
    &\nabla \cdot \vB
        = 0 \,
\qquad\qquad\quad \mbox{and} \qquad \qquad\quad
    \nabla\times \vB
        =   4\,\pi\,\frac{\vj}{c}
               + \frac{1}{c}\,\frac{\partial \vE}{\partial t}
               + \nabla q\,.
\end{aligned}
\end{equation}
Here $\rho$ and $\vj$ differ from their classical counterparts
and the new laws of Gauss and Ampere differ from their
classical counterparts, too (cf.~\cite{Fey89II, LanLif89,Jac99}).
Moreover, we see that Maxwell's equations~(\ref{Maxwell01b}) depend not
only on the fields $\vE$ and $\vB$ but also on the scalar
$q=\partial_\mu\,A^\mu$.

\section{Correspondence principle}
\label{sec-Cp1}

Now it is time to  show how classical electrodynamics can be obtained via 
a special ``time averaging process'' from our non-instant field model. 
As in quantum mechanics we call this the \emph{correspondence principle} 
of our field model. At the end of this section we examine the 
\emph{Aharonov-Bohm effect} from the point of view of the non-instant 
field model together with the correspondence principle.

\subsection{Covariant time averaging}

For an introduction into the principle of covariance we refer to~\cite{Wei72}. 
We define the covariant time average of the non-instant four potential
$A^\mu$ by
\begin{equation}\label{averaging}
\begin{aligned}
     \A[A^\mu](\x,T)
      := \int_{T}^{T+\tau} A^\mu(\x,t) \,\frac{\d t}{\tau}\,,
\end{aligned}
\end{equation}
where $\tau =\tau'/\sqrt{1-v^2/c^2}$ and $\tau'$ is defined as in
Subsection~\ref{sec-defA}.
This is a four vector, because
\begin{itemize}
\item $A^\mu$ is a four vector,

\item $\d t\,\sqrt{1-v^2/c^2}$ and $\tau\,\sqrt{1-v^2/c^2}$ are scalars.

\end{itemize}
The time averaging of other tensors is defined analogously. \\

\begin{rema}\label{rema-opA}
We note that in general $\A[X(t)\,Y(t)] \not = \A[X(t)]\,\A[Y(t)]$ 
for \emph{two time-dependent} quantities $X(t)$ and $Y(t)$. 
In particular,
$$
       \A[X(t)]=0
\qquad \not\Rightarrow \qquad
          \A[X(t)\,Y(t)]=0\,.
$$
For example, this fact is important for the calculation of the 
energy content of ${A'}_j^\mu$ in Section~\ref{sec-energy}.
\end{rema}

\subsection{Derivation of the classical equations of Maxwell}

In the following we assume that all quantities are sufficiently smooth.
From~(\ref{averaging}), it follows at once that
\begin{equation}\label{propAA01}
    \frac{\partial \A[A^\mu]}{\partial x_j}(\x,t)
       = \A\left[\frac{\partial A^\mu}{\partial x_j}\right](\x,t)
\qquad\quad j=1,\,2,\,3\,.
\end{equation}
Moreover, from analysis it is known that (cf.~\cite{Lan83})
\begin{equation*}
\begin{aligned}
   \frac{\partial }{\partial T}\,\int_{T}^{T+\tau} g(t)\,\frac{\d t}{\tau}
    = \frac{g(T+\tau) - g(T)}{\tau}
    = \int_{T}^{T+\tau} \frac{\partial g(t)}{\partial t}\,\frac{\d t}{\tau}
\end{aligned}
\end{equation*}
and therefore we have
\begin{equation}\label{propAA02}
  \frac{\partial \A[A^\mu]}{\partial t}(\x,t)
    = \A\left[\frac{\partial A^\mu}{\partial t}\right](\x,t)\,.
\end{equation}
Similar statements hold for other tensors. Hence the application of the 
operator $\A$ onto the equations of Maxwell~(\ref{Maxwell01a}) yields
\begin{equation*}
\begin{aligned}
  \partial_\nu \,\A[F^{\mu\,\nu}] - \partial^\mu \,\A[q]
   =  - 4\,\pi\,\frac{\A[j^\mu]}{c}\,.
\end{aligned}
\end{equation*}
In the following subsection, we show that $\A[q]=0$ holds for a
system of charged particles, i.e. the Lorentz condition is satisfied
for the time averaged field.

\subsection{Classical fields of a charged particle}
\label{sec-classfield}

For simplicity  we assume that the restframe $S'$ of the charged particle 
$Q_0$ is an inertial system such that ${A'}^\mu = \sum_{j\in J} {A'}^\mu_j$ 
holds in $S'$ (cf.~(\ref{Atotal})). Moreover, we assume that the particle 
lives in $S'$ for $t\in (-\infty ,\infty)$, i.e. $J=\Z$.

In the restframe $S'$ of the charged particle, the electric and magnetic 
fields are
\begin{equation}\label{EA}
\begin{aligned}
  \vE'(\x',t')
     &= \frac{Q_0}{c}\,\sum_{j\in\Z} \frac{\frac{\d f'_0}{\d t'}
                           (\omega'\,(t'-\tau_j')-k'\,|\x'|)}{|\x'|}\,\e_{R'} \\
 & \qquad+ Q_0\,\sum_{j\in\Z} \frac{f'_0(\omega'\,(t'-\tau_j')-k'\,|\x'|)}
                                 {|\x'|^2} \,\e_{R'},\\
  \vB'(\x',t')
     &= 0\,,
\end{aligned}
\end{equation}
where $\e_R':=\nabla'\,|\x'|$. From this,~(\ref{defA}),~(\ref{deff0}) and
$\A[\frac{\partial f'_0}{\partial t'}]=0$, we get
\begin{equation*}
\begin{aligned}
   &\A[\phi'](\x',t')
       =  \frac{Q_0}{|\x'|}\,,
\qquad\qquad\qquad\qquad\quad\;
   \A[\vA'](\x',t')=\vO \,, \\
    &\A[\vE'](\x',t')
         = -\nabla' \A[\phi'](\x')
\qquad  \mbox{and}  \qquad
   \A[\vB'](\x',t')=\vO\,,
\end{aligned}
\end{equation*}
which are nothing else but the quantities of classical electrodynamics. 
From these equations,~(\ref{propAA01}),~(\ref{propAA02}) and~(\ref{defj}), it 
follows that
\begin{equation*}
\begin{aligned}
    \A[q'] = \nabla'\cdot \A[\vA']
         + \frac{1}{c}\,\frac{\partial \A[\phi']}{\partial t}
          = 0
\qquad  \mbox{and}  \qquad
     \A[j'^\mu]:=(c\,Q_0\,\delta(\x),\vO) \,.
\end{aligned}
\end{equation*}
Here we have used that $\A[\vA']=\vO$ and the classical field $\A[\phi']$ does 
not depend on time. Thus we interprete $\A[A^\mu]$ as the (ideal) data of the four 
potential $A^\mu$ obtained by a real measurement.\\

\begin{rema}\label{rema-AhaBoh}
Assume for the moment that the exact values of $A^\mu$ cannot be measured by 
experiment. Does this mean that it is not reasonable to use the non-instant 
field $A^\mu$ in calculations? If the quantity $A^\mu$ is related to reality 
it must have an effect. The effect is the warranty of causality, i.e. using 
only $\A[A^\mu]$ yields effects without cause, like the Aharonov-Bohm effect 
(cf.~\cite{AhaBoh59,Fey89II,Sch93,Gre93,Zee03}) but using $A^\mu$ gives us 
the cause. This will be shown in the following two subsections.\\
\end{rema}

\subsection{The field of a solenoid}

\begin{figure}[!ht]
\begin{center}
\includegraphics[height=6.0cm,angle=0]{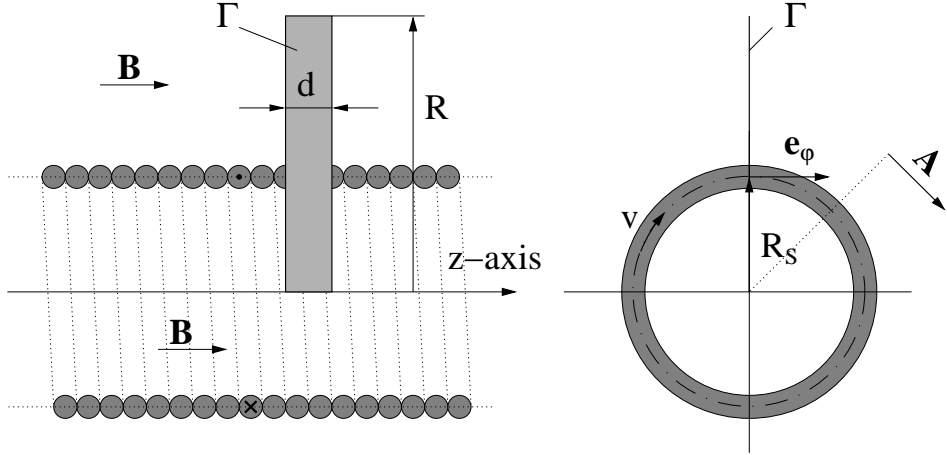}
\end{center}
\caption{Visualization of a solenoid and its magnetic field.}
\label{fig:ampere}
\end{figure}

Consider a solenoid with axis $\{(0,0,z)\,|\, z\in\R\}$ and radius $R_S>0$
that was switched on at time $t=-\infty$ (cf. Fig.~\ref{fig:ampere}).
Our reference frame is the restframe of the solenoid. Because the speed
$v$ of the electrons in the solenoid are much smaller than the speed
$c$ of light, we model the current in the solenoid by 
\begin{equation}\label{Solj} 
\begin{aligned}
   \vj = \sum_{i\in I} \sigma_i(t)\,v\,\frac{\delta(R-R_S)}{R}\,\e_\varphi\,,
\end{aligned}
\end{equation}
with the assumptions 
$$
         \A[\sigma_i] = \sigma=const. 
\qquad\mbox{and}\qquad
          v=const. 
$$
For a fixed $i\in I$  $\sigma_i=\sigma_i(t)$ denotes the surface charge
density of the electrons that send out their fields at the same time. 
In the ``time averaged picture'' we have the 
time-independent source term 
\begin{equation}\label{SolAj}
\begin{aligned}
      \A[\vj] = \sigma\,v\,\frac{\delta(R-R_S)}{R}\,\e_\varphi \,.
\end{aligned}
\end{equation}
Here the important point is that in contrast to~(\ref{SolAj}), the "real'' 
source term~(\ref{Solj}) is time-dependent. 

The symmetry of the solenoid implies $\vB(\x,t)=B(|\x|,t)\,\e_z$ which
together with $\vB=\nabla\times \vA$ implies
$$
   \vA(\x,t) = A(|\x|,t)\,\e_\varphi \,.
$$
For the solution of~(\ref{Maxwell01}) with time-dependent source 
term~(\ref{Solj}), it follows from pde theory (cf.~\cite{Tre95}) that 
$\vA$ does not vanish and is time-dependent outside the solenoid and 
therefore 
$$
     \vE_w:=-\frac{1}{c}\,\frac{\partial \vA}{\partial t}  \qquad \mbox{(not identical zero)}
$$
implies due to the law of induction 
$\nabla\times \vE_w = - \frac{1}{c}\,\frac{\partial \vB_w}{\partial t}$
a non-vanishing magnetic field
$$
     \vB_w = \nabla \times \vA\,,
$$
which is orthogonal to $\vE_w$.
Indeed, $(\vE_w,\vB_w)$ is a wave with extrem high frequency $2\,\pi/\tau$ 
(cf.~(\ref{defA}) and~(\ref{defomegak})) such that
$$
   \A[\vE_w]
     = 0
\quad\mbox{and}\quad
   \A[\vB_w] = 0 \qquad \mbox{outside the solenoid.}
$$
This can be seen as follows. The time-independence of $\A[\vj]$ implies 
the time-independence of $\A[\vA]$ and consequently by~(\ref{propAA02})  
$$
     \A[\vE_w] 
     = -\frac{1}{c}\,\frac{\partial \A[\vA]}{\partial t}
     = 0\,. 
$$ 
This and the second equation in~(\ref{Maxwell01b}) together with~(\ref{propAA01}) 
and~(\ref{propAA02}) imply 
$$
  - \frac{1}{c}\,\frac{\partial \A[\vB_w]}{\partial t} 
     = \nabla\times \A[\vE_w]
     =  0 \,
$$
and consequently $\A[\vB_w]$ is a constant. That $\A[\vB_w]$ vanishes 
follows from the fact that $\A[\vB_w]$ vanishes at infinite. 

Since it is very instructive, we give another prove of $\A[\vB_w] =0$ 
via the law of Ampere (cf.~\cite{Fey89II}). Consider the situation 
visualized in Fig.~\ref{fig:ampere}. 
From the law of Ampere (cf. last equation in~(\ref{Maxwell01b})), the 
theorem of Stokes (cf.~\cite{Lan83}) and the symmetry of the problem, 
it follows
\begin{equation*}
\begin{aligned}
    (B(0,t) -B(R,t))\,d
         &= \int_{\partial \Gamma} \vB\cdot \d \x
         = \int_{\Gamma} (\nabla\times \vB)\cdot \e_\varphi \,\d S \\
         &= 4\,\pi\,\int_{\Gamma}\left (\frac{j}{c}+\nabla q\right)\,\d S
            + \frac{1}{c}\,\int_{\Gamma} \frac{\partial\vE}{\partial t}\,\d S\,.
\end{aligned}
\end{equation*}
Because of $\A[q]=0$ and $\A[\frac{\partial \vE}{\partial t}]=0$ (due to 
$\A[j^\mu]=0$), application of the time averaging operator $\A$ to our 
last result yields
\begin{equation}\label{AverAmper}
\begin{aligned}
 \A[B](0,t) - \A[B](R,t)
  = \frac{4\,\pi}{c\,d}\,
               \int_{\Gamma} \A[j]\,\d S \,.
\end{aligned}
\end{equation}
Here we have used the assumption that the solenoid was switched on at time 
$t=-\infty$. But~(\ref{AverAmper}) implies at once $\A[B](R_1,t) = \A[B](R_2,t)$ 
for all $R_1,\,R_2>R_S$. Because $\A[B](R,t)\stackrel{R\to\infty}{\longrightarrow}0$ 
must hold, we infer
\begin{equation}\label{solAB}
\begin{aligned}
    \A[B](R,t) = 0 \qquad\mbox{for all}\qquad R>R_S\,,
\end{aligned}
\end{equation}
as was to be shown. This calculation shows in which way classical electrodynamics 
fits into our non-instant field theory.

\vspace{0.5cm}
In summary, the non-instant field model implies that a solenoid emits a wave with 
extrem high frequency $2\,\pi/\tau$ and since the time interval $\tau$ 
is extremely small, a measurement of the electromagnetic field outside the solenoid 
yields the mean zero (correspondence principle).

\subsection{The Aharonov-Bohm effect}

In this subsection we show that the Aharonov-Bohm effect does not violate causality 
if the non-instant field model is used and that the classical result follows after 
time averaging.

\begin{figure}[!ht]
\begin{center}
\includegraphics[height=6.5cm,angle=0]{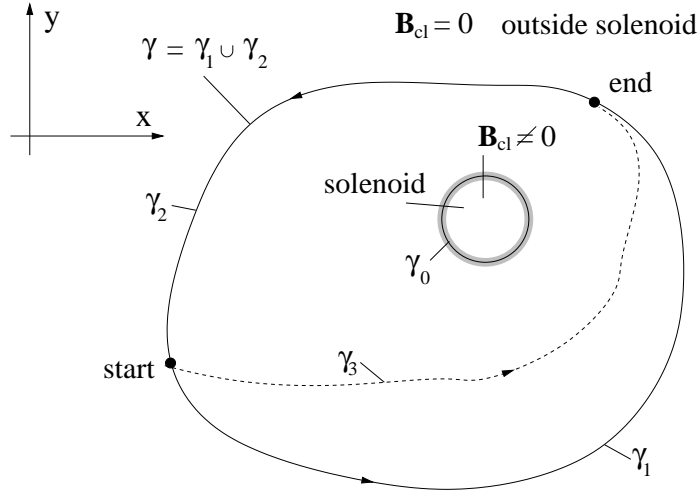}
\end{center}
\caption{A consequence of the Aharonov-Bohm effect is that 
$\int_{\gamma} \vA_{cl}\cdot \d \x=\int_{\gamma_0} \vA_{cl}\cdot \d \x$ 
for each $\gamma$ enclosing $\gamma_0$. In particular,  
$\int_{\gamma_1} \vA_{cl}\cdot \d \x=\int_{\gamma_3} \vA_{cl}\cdot \d \x$.}
\label{fig:bohm}
\end{figure}

First we explain the classical point of view of the Aharonov-Bohm effect. Although 
measurements show that there is no magnetic field outside a solenoid, an electron 
passing by is effected if the solenoid was switched on. In mathematical terms this 
means that due to Stokes theorem and $B_{cl}\,\e_z=\nabla\times \vA_{cl}$, the classical 
quantities $\vA_{cl}$  and $\vB_{cl}$, which we identify as $\A[\vA]$  and $\A[\vB]$, 
satisfy (cf. Fig.~\ref{fig:bohm})
\begin{equation}\label{noncausalBohm}
\begin{aligned}
    \int_{\gamma} \vA_{cl}\cdot \d \x
       = \int_{\Gamma} \vB_{cl}\cdot \d \vS
       = \int_{\Gamma_0} \vB_{cl}\cdot\d \vS   
       =\int_{\gamma_0} \vA_{cl}\cdot \d \x 
\end{aligned}
\end{equation}
for every
$$
   \mbox{closed path $\gamma$ that encloses the simple curve $\gamma_0$.}
$$ 
Here the sets $\Gamma$ and $\Gamma_0$ are the closures of $\gamma$ and $\gamma_0$, 
respectively. Thus it appears as if ``information'' is transported infinitely fast 
from the solenoid to the curve $\gamma=\partial\Gamma$ enclosing the solenoid. 
In other words, the electron passing by feels somehow the magnetic field inside 
the solenoid and its motion is influence by it. This is the enigma of the 
Aharonov-Bohm effect. 

What happens if we use the non-instant field model? We already know from the 
previous subsection that there is a wave which transports information with 
the speed of light $c$. Let's check the mathematics. From $\vB=B\,\e_z$ and the 
theorem of Stokes we get
\begin{equation}\label{causalBohm}
\begin{aligned}
    \int_{\partial\Gamma} \vA\cdot \d \x
       = \int_{\Gamma} (\nabla\times\vA)\cdot \e_z\,\d S
       = \int_{\Gamma} \vB\cdot\d \vS  \,       
       = \int_{\Gamma_0} \vB\cdot\d \vS 
         + \int_{\Gamma\backslash\Gamma_0} \vB\cdot\d \vS 
\end{aligned}
\end{equation}
in which causality is not violated. Due to $\A[\vB]=0$ outside the 
solenoid (cf.~(\ref{solAB})), the last term satisfies the property 
\begin{equation*}
\begin{aligned}
         \int_{\Gamma\backslash\Gamma_0} \A[\vB]\cdot\d \vS = 0 \,.
\end{aligned}
\end{equation*} 
From this, the assumption that the solenoid was switched on at time 
$t=-\infty$ and~(\ref{propAA01}), it follows after time averaging 
of~(\ref{causalBohm}) that 
\begin{equation*}
\begin{aligned}
    \int_{\partial\Gamma} \A[\vA]\cdot \d \x
       = \int_{\Gamma_0} \A[\vB]\cdot\d \vS 
       =  \int_{\partial\Gamma_0} \A[\vA]\cdot \d \x\,,
\end{aligned}
\end{equation*}
which is according to our correspondence principle nothing else 
but~(\ref{noncausalBohm}). Thus if $\vB$ is considered as the real 
magnetic field, then the information is transported with the speed 
of light but not faster. More to the point, it is causality that 
``guarantees'' the Aharonov-Bohm effect and permits - in the time 
averaged picture - a Lorentz gauge.

\section{Conservation of energy and the quantization of field energy}
\label{sec-energy}

\subsection{The conservation law}

The energy conservation law of the non-instant field model reads as follows
\begin{equation}\label{Econserv}
\begin{aligned}
  \nabla \cdot \vS + \frac{1}{c}\,\frac{\partial u}{\partial t}
       = 4\,\pi\,\left( \rho\,q -\frac{\vj}{c}\cdot \vE
                  \right) \,,
\end{aligned}
\end{equation}
where $q=\partial_\mu A^\mu$ and 
\begin{equation}\label{defuS}
     u := \frac{1}{2}\,(q^2 + |\vE|^2 + |\vB|^2)
\qquad \mbox{and}\qquad
     \vS := q\,\vE + \vE\times \vB \,
\end{equation}
correspond to the \emph{energy density} and the \emph{energy flux} 
(\emph{Poynting vector}) of the electromagnetic field, respectively. 
Moreover, $j^\mu=(c\,\rho,\vj)$ denotes the four current and the right 
hand side models the source of energy. \\

This conservation law can be derived similarly as in classical field theory 
(cf.~\cite{LanLif89}): Scalar multiplication of the second equation 
in~(\ref{Maxwell01b}) with $\vB$ and the last equation in~(\ref{Maxwell01b}) 
with $\vE$ together with identity
$$
    \nabla\cdot (\vE\times \vB)
      = \vB\cdot (\nabla \times\vE)
        - \vE\cdot (\nabla \times\vB)
$$
yields
\begin{equation}\label{helpEnergy}
\begin{aligned}
  \frac{1}{c}\,\frac{\partial}{\partial t} \frac{|\vE|^2+|\vB|^2}{2}
         + 4\,\pi\,\frac{\vj}{c}\cdot \vE + \nabla\cdot( q\,\vE) -q\,\nabla \cdot \vE
         + \nabla \cdot (\vE\times \vB)  = 0\,. 
\end{aligned}
\end{equation}
Because of the law of Gauss (cf.~(\ref{Maxwell01b})) 
\begin{equation*}
\begin{aligned}
    &\nabla \cdot \vE
        = 4\,\pi\,\rho
         - \frac{1}{c}\,\frac{\partial q}{\partial t}\,,
\end{aligned}
\end{equation*}
equation~(\ref{helpEnergy}) is equivalent to the \emph{energy conservation law}~(\ref{Econserv}) 
with $u$ and $\vS$ defined as in~(\ref{defuS}).

\subsection{Quantized energy of the field ${A'}_j^\mu$}

In the following we consider a single electron $Q_0:=e^-$ that has life time 
$N\,\tau'$ ($N\in\N$) in its restframe $S'$. We show that the energy $E_j'$ 
of the field ${A'}_j^\mu$ in $S_j'$ is given by
\begin{equation}\label{energyEpj}
\begin{aligned}
  E'_j = \omega'\,\hbar   \qquad\mbox{where}\qquad 
  \omega' = \frac{2\,\pi}{\tau'}\,.
\end{aligned}
\end{equation}
We recall that $S'_j$ as the inertial systems such that its origin and axes 
coincide with those of the restframe $S'$ of the electron at time $t'=\tau_j'$ 
when the field is sent out (cf. Subsection~\ref{sec-defA}). And thus the total 
energy sent out by the electron in $S'$ during its life time $N\,\tau'$ is
$$
    E'_{total} = N\, \omega'\,\hbar\,.
$$
Due to the quantization of charges, it follows that the field energy ${A'}_j^\mu$ 
of a charged particle is quantized. \\

Since $E'_j=E'_0$ for all $j\in J$ we restrict the following calculating to the 
index $j=0$. In the frame $S_0'$ we have $\vj=\vO$ (vector current) and $\vB'=\vO$, 
thus the conservation law simplifies to
\begin{equation}\label{conteq2}
\begin{aligned}
 \frac{\partial u'_0}{\partial t'}
         + c\,\nabla' \cdot \vS_0'
       = 4\,\pi\,\rho_0'\, q_0'
\qquad \mbox{with}\qquad
   \vS_0' = q_0'\, \vE'_0\,
\end{aligned}
\end{equation}
and $q_0'$, $\rho_0'$ defined as in~(\ref{defqj}),~(\ref{defj}). From this and the 
divergence theorem (cf.~\cite{Lan83}), it follows that the energy $E_0'$ 
is given by
\begin{equation*}
\begin{aligned}
  E'_0
      = c\, \int_{0}^{\tau'} \int_{B_{R'}(\vO)} (\nabla' \cdot \vS_0')
                  \,\d V'\,\d t'
      = c\,\tau'\,\int_{\partial B_{R'}(\vO)} \A[\vS_0'] \cdot \e_{R'}\,\d S'
      \quad \mbox{with}\quad  R'=c\,\tau'.
\end{aligned}
\end{equation*}
In classical electrodynamics we have $\vS_0'=\vO$, but because of
$$
    \A[q_0'\,\vE'_0] \not= \A[q'_0]\,\A[\vE'_0] = 0   \qquad\quad (\A[q'_0]=0)\,,
$$
this is not true for the non-instant field model. From 
$q_0'=\frac{1}{c}\,\frac{\partial \phi_0'}{\partial t'}$ with~(\ref{defA}),~(\ref{defomegak}),~(\ref{EA}) 
and~(\ref{deff0}), it follows that
$$
  \A[\vS_0'](\x',t')\cdot \e_{R'}
    =  \frac{e^2}{\tau'\,c^2\,|\x'|^2}\,
      \int_\R\left[\left(\frac{\partial f'_0(\omega'\,t')}
                          {\partial t'}\right)^2\right] \,\d t'
    = \frac{\omega'\,\hbar}{4\,\pi\,\tau'\,c\,|\x'|^2} \,,
$$
since
$$
   \A \left[  \frac{\partial f'_0}{\partial t'}(\omega'\,t')\,
          f'_0(\omega'\,t') \right] =0
\quad\mbox{and}\quad
   \int_\R\left[\left(\frac{\partial f'_0(\omega'\,t')}
                     {\partial t'}\right)^2 \right]\,\d t'
     =   \frac{\omega'\,c\,\hbar}{4\,\pi\,e^2}\,.
$$
From this we get the energy content of ${A_0'}^\mu$ as
\begin{equation*}
\begin{aligned}
  E'_0
      = \int_{\partial B_{R'}(\vO)} \frac{\omega'\,\hbar}{4\,\pi\,|\x'|^2}\,\d S'
      = \omega'\,\hbar \qquad\quad  (R'=c\,\tau')\,.
\end{aligned}
\end{equation*}
This proves our claim~(\ref{energyEpj}) for the 
field ${A'}_j^\mu$ for $j\in J$. \\

\subsection{Is there an energy problem?}

As mentioned in Remark~\ref{rema-opA}, we have 
$$
           \A[X(t)\,Y(t)] \not= \A[X(t)]\,\A[Y(t)]\,
$$
and therefore the correspondence principle cannot be applied to products 
of physical quantities.  Let us look closer at this in the context of field 
energy. Let $u_{cl}$ and $\vS_{cl}$ denote the energy density and Poynting 
vector used in classical field theory, i.e
\begin{equation*}
     u_{cl} := \frac{1}{2}\,(\A[E]^2 + \A[B]^2) = \A[u_{cl}]
\quad \mbox{and}\quad
     \vS_{cl} := \A[\vE]\times \A[\vB] = \A[\vS_{cl}]\,.
\end{equation*}
Because of
$$
       \A[{q'}^2] \not= \A[q']\,\A[q'] = 0
       \qquad\mbox{and}\qquad
       \A[q'\,\vE'] \not= \A[q']\,\A[\vE'] = 0\,,
$$
it follows that
$$
       \A[u'] \not= \A[u_{cl}']
       \qquad\mbox{and}\qquad
       \A[\vS'] \not= \A[\vS_{cl}'] = 0
$$
in the restframe of the charge. Consequently,
$$
       E_j' = \omega'\,\hbar \not= 0 =
       c\,\tau'\,\int_{\partial B_{R'}(\vO)} \A[\vS_{cl}'] \cdot \e_{R'}\,\d S \,,
$$
i.e. according to the non-instant field model, a charge sends out quantized energy, 
which is in contrast to the standard model. But this has to be expected 
of a non-instant field model.

As a consequence we have to ask: "Is there an energy problem for charged 
particles?" My point of view is that a charged particle is able to use the 
energy from its surrounding (e.g. the microwave background radiation) to 
maintain its field. Besides, there is plenty of energy near our sun. 
In this way energy is spread over the universe and can 
be used by other particles. (This is very suggestive, is it not?) Geometric 
considerations imply that a  macroscopic object can maintain its required 
energy content much longer than a single particle. 
The worst case is a single particle far away from energy suppliers. So, 
how does a charged particle react if it does not get enough energy? Is it
\begin{itemize}
\item destroyed, or

\item does it use its rest mass (or a part of it) to maintain its field or

\item does it move without sending out an electromagnetic field?
\end{itemize}
I consider these questions as very intriguing, but it is beyond the scope 
of this paper to clarify and solve them.

\section{Results}

We have presented a non-instant field model for classical field theory that 
can be quantized with the \emph{Gupta-Bleuler method}. The fact that the 
non-instant field model does not permit a Lorentz gauge transformation is 
not an obstacle for quantization and does not contradict the essential results 
of \emph{classical field theory}. 
Indeed, our model obeys a correspondence principle, i.e. the classical field 
quantities $\A_{cl}^\mu$, $\vE_{cl}$ and $\vB_{cl}$  can be obtained via a 
covariant time averaging operator $\A$ from the respective quantities of the 
non-instant field model and the Lorentz gauge transformation is satisfied for 
the time averaged four potential. 
Although the calculation with the time averaged quantities is very convenient, 
it does not allow to associate a cause to each effect. 
As shown in this paper, the non-instant field model predicts that a solenoid 
sends out an electromagnetic wave of extreme high frequency such that
$\A[\vE]=\vO$ and $\A[\vB]=\vO$ outside the solenoid, 
and consequently - according to our correspondence principle - a measurement of 
the magnetic field yields the value zero outside the solenoid. As a consequence, 
the \emph{Aharonov-Bohm effect} does not violate causality if the non-instant field 
model is used. However, if we go over to the "time averaged picture" ($\equiv$ 
classical electrodynamics), causality seems to be violated.

In addition, we have shown that the field energy of the non-instant field model is 
quantized.

\subsection*{Conclusions}

As we have demonstrated, classical field theory describes the 
behavior of time averaged field quantities $\A[X],\,\A[Y],\,\ldots$ rather 
than the non-instant field quantities $X,\,Y,\,\ldots$ themselves. If we 
grant the quantities $X,\,Y,\,\ldots$ physical significance by using a 
non-instant field theory, then we are able to describe the \emph{Aharonov-Bohm effect} 
in a causal way and the quantization of the respective equations of Maxwell 
is still possible. If you find gauge theories ``deeply disturbing and unsatisfactory'' 
(cf. Chapter III.4 in~\cite{Zee03}), then the non-instant model for $X,\,Y,\,\ldots$ 
is the right thing, but if you like gauge theories, then you go over to the time 
averaged picture $\A[X],\,\A[Y],\,\ldots$. In the latter case, non-causality is 
not always meaningful. 
The role of the covariant time averaging operator $\A$ 
is understood as a causal restriction on the measurement process. That is 
to say, every measurement of a quantity $X$ will require at least a minimal 
time period $\tau$ and produces the time averaged value $\A[X]$. (Of course 
this is an idealized picture and the time period $\tau$ depends on the 
observers reference frame.) In other word, even the most cunning experimental 
setup will not permit to measure $X$ and $Y$, but $\A[X]$ and $\A[Y]$, and if 
$$
     X\not=Y  \qquad \mbox{and}\qquad \A[X]=\A[Y],
$$
then there will be an experimental setup that shows $X\not=Y$. However, if you 
misinterpret your result, then you conclude a failure of logic or causality.

\end{document}